\documentstyle[PASJadd,psfig]{PASJ95}
\newcommand{\vol}{52}
\newcommand{\no}{5}
\newcommand{\lett}{}
\newcommand{\spage}{}
\newcommand{\rdate}{2000 January 20}
\newcommand{\adate}{2000 July 10}
\begin{document}
\title{%
Profiles and Polarization Properties of 
Emission Lines from Relativistic  Disks
 }

\author{%
  Jun {\sc Ogura},
  Nobuyori {\sc Ohuo}
  \thanks{Present address: Institute for Hydrospheric-Atmospheric
  Sciences, Nagoya Univ., Nagoya 464--8601.},
  and
  Yasufumi {\sc Kojima} \\
  \it{%
  Department of Physics, Hiroshima University, Higashi-Hiroshima
  739--8526
  }\\
  {\it E-mail(JO):ogura@theo.phys.sci.hiroshima-u.ac.jp}
}

\abst{
We examined the profiles and polarization properties of emission lines 
originating from the extremely relativistic region of Kepler disks.
Because the emission region is more localized to the inner part,
e.g., a few times the black hole radius,
the profiles, and polarization properties are remarkably modified. 
The black hole spin significantly affects the properties 
through the inner edge.
This observation, in spite of the technical difficulties,
provides an important diagnostic of 
the relativistic region.   
}

\kword{black hole physics --- line: profiles --- polarization}

\maketitle
\thispagestyle{headings}
%%%%%%%%%%%%                Section  1             %%%%%%%%%%%%%%%%%%%
%
%section 1
%
\section{Introduction}

%1
  In recent years, the evidence of black holes has become more apparent 
owing to X-ray astronomy. Some broad iron lines from Seyfert 
galaxies were detected and attributed to relativistic effects 
in the emission region (Tanaka et al. 1995; Iwasawa et al. 1996; Nandra
et al. 1997).
The lines emitted from accretion disks near the central black hole show
broad asymmetric profiles, significantly 
modified by both the gravitational redshift and Doppler effects. 
The emission region is inferred to be located near several 
times the black hole horizon, since the line width 
determines the strength of the relativistic effect, i.e., 
the radius from the disk center.  Strong gravitational 
effects are crucial there, and the angular momentum of a rotating black
hole is also constrained. (Dabrowski et al. 1997; Weaver, Yaqoob 1998)

%2
  The theoretical possibility was discussed beforehand.
Fabian et al. (1989) calculated the line profile   
emitted from accretion disks around non-rotating black holes.
Kojima (1991) and Laor (1991) independently
extended this study to rotating cases.
The effect of black hole rotation appears only in the disk
structure close to the central black hole.
The present remarkable observation gives not only evidence, but also
the property of the central black holes.
Future observations will give new topics
related to strong gravitational phenomena.

%3
Some natures of central black holes, which will be observable by
the next advanced technique, are also discussed.  
Reynolds et al. (1999) calculated an X-ray reverberation map
of lines for flares on the disk, which 
constrains the mass and spin of the black hole.
In addition to the intensity and  its time variation, 
the polarization of light is important. 
Polarization is expected to be produced by electron
scattering above the accretion disk. 
Several authors (Connors et al. 1980: Laor et al. 1990)
have discussed the possibility.
In particular, polarization in the emission line may be important,
although the observational technique is very difficult.
The emission region of the line is inferred by the profile, 
so that the polarization is an additional diagnostic tool.
Chen and Eardley (1991) already studied the polarization
properties of the line originating from a disk around 
a Schwarzschild black hole.
It is important to extend their calculation 
to a Kerr black hole and to explore the  effect of 
the black hole rotation, since 
the current observation of the broad lines suggests
likely rotating black holes.
In this paper,  we discuss a theoretical 
calculation of the possibility in advance, that is,
the polarization and line profile originating from 
a disk around a Kerr black hole.
%4 organization
In the next section, we describe our models.
The line profiles and the polarizations depend on several parameters. 
In section 3, the numerical results are given to demonstrate typical
examples.
We show how additional information can be
useful for determining the models.
Section 4 is devoted to a discussion. 
We use the units $c=G=1$ throughout the paper.

%
%section 2
%
\section{Assumptions and Calculation Methods}
%\label{chap:a-c}

%0
In this section, we briefly summarize our model and
the numerical method. The details of the formalism are given elsewhere
e.g., for the line profile (Kojima 1991) 
and the polarization (Connors et al. 1980).

%1
   The polarization of a beam of radiation can be 
described by the Stokes parameters (Chandrasekhar 1960),
i.e., the intensity $I$, linear polarization parameters $Q$, $U$,
and the circular parameter $V$. The circular parameter
vanishes for polarization induced by scattering.
These quantities are related to the degree of
polarization, $\delta$, and the angle 
of the plane of the polarization, $\psi$, as 
%
%  equation 1
\begin{equation}
\delta=\frac{\sqrt{Q^{2}+U^{2}} }{I },
\hspace{5mm} 
\psi=\frac{1}{2}\tan^{-1}\left(\frac{U}{Q}\right).
\end{equation}
%

%2 Emission

  For a purely electron-scattering atmosphere,
the polarization is determined by the local physics. 
The degree of polarization depends on the  
emission angle, $\phi_{{\rm e}}$, which is measured in the
rest frame.  
An explicit formula has been derived by Chandrasekhar (1960).
The degree, $\delta_{{\rm e}}$, ranges from 12\% for $\phi_{{\rm
e}}=90^{\circ}$ (edge-on case) 
to 0 for $\phi_{{\rm e}}=0$ (face-on case).
%

%3 Propagation
These polarization  quantities as well as the energy, $E_{{\rm em}}$, in
the rest frame of emitting matter are 
related to the observed ones by the  null geodesics.
The observed energy, $ E_{{\rm ob}} $, is shifted as 
$E_{{\rm ob}} =g E_{{\rm em}}$, with the redshift  factor  $ g^{-1}. $
Since the polarization degree, $\delta$, is a scalar,
we have the observed value as  
$\delta = \delta_{{\rm e}} $, for the corresponding path.
  The angle of the  polarization plane measured by an observer 
can be calculated by the parallel transportation along the null 
geodesics. 
The polarization vector at infinity can be constructed
with the help of the so-called Walker--Penrose constant in
the Kerr metric. 

%model

In this paper, we assume  a geometrically thin, axisymmetric 
disk co-rotating in the equatorial plane around 
a Kerr black hole.
The polarization is assumed to be produced  by electron 
scattering in the disk surface
ranging from $R_{{\rm in}}$ to $R_{{\rm out}}$.
We adopt a power law-form with a constant index, $q$,
as the local specific intensity 
$\Bigl[I_{{\rm e}}(E_{{\rm em}}, \phi_{{\rm e}}, r_{{\rm e}})\Bigr]$;
% equation 2
\begin{equation}
I_{{\rm e}}(E_{{\rm em}}, \phi_{{\rm e}}, r_{{\rm e}}) \propto
\frac{\epsilon(\phi_{{\rm e}})}{4\pi}r^{-q} \delta(E_{{\rm em}} -E_{{\rm o}}),
\label{s-intensity}
\end{equation}
where $E_{{\rm o}}$ is the rest energy of the line and
$\epsilon(\phi_{{\rm e}}) $ describes the local 
line specific intensity.
We assumed that the intrinsic broadening of the line 
is so small that it can be neglected compared to the kinematical effect.

The flux  by an observer at $\theta_{{\rm o}}$
is calculated by integrating over the disk surface,
%
% equation 3
\begin{equation}
  F_{{\rm o}}(E,\theta_{{\rm o}})
            =\int \int g^3 I_{{\rm e}} (E/g,\phi_{{\rm e}},r_{{\rm e}})
{\rm d} \Omega .
\end{equation}
In a similar way, 
the observed net Stokes parameters $\bar{Q},\bar{U}$ are
calculated as
% equation 4
\begin{eqnarray}
&\bar{Q}_{{\rm o}}(E,\theta_{{\rm o}})
                       +i \bar{U}_{{\rm o}}(E,\theta_{{\rm o}})
  \hspace*{3.5cm}& \nonumber\\
 &=
\int \int g^3 \delta_{\rm e} ( \cos 2 \psi  +i \sin 2 \psi )
I_{\rm e}(E/g,\phi_{{\rm e}},r_{{\rm e}})
{\rm d} \Omega&
\end{eqnarray}
The observed degree, $\delta$, and 
angle of polarization, $\psi$, are calculated by equation (1).
The angle is measured from the direction perpendicular to
both the normal vector of the disk and the propagation vector
of the light.
These quantities depend on the disk inclination, $ \theta _{\rm o}$, the model
parameters 
$q, R_{{\rm in}}, R_{{\rm out}},$ and the Kerr parameter $a$.
It is possible to calculate them in this large
parameter space, but results are not easy to be understood.
Since we found that the dependence of $q$ is not very sensitive, 
the results for $q=3$ are shown in the next section.
As the outer boundary, $R_{{\rm out}}$, of the disk increases,
the relativistic effects are less prominent because of the 
average over the disk. We will show a  representative result 
of $R_{{\rm out}}=20 M$ in the next section. Here $M$ is the mass of
the central black hole.

%\newpage
%
% section 3
%
\section{Numerical Results}

%1
Before discussing the realistic cases, we illustrate the polarization
angles emitted from a few rings in figure 1.
Because the general cases can be regarded as a summation over a
certain range of the disk, this figure is useful for
understanding the relativistic effects.
The shift in the energy and the polarization angle 
is small for emission from a ring of large radius. 
As the radius decreases,
the range of the observed energy is more spread and the
polarization angle significantly deviates from $ 90^{\circ}. $
These quantities strongly depend on the emission position,
and therefore give the information about the emission region
as well as the line profile.

%2 : Fig.2(a)(b)

In figure 2-a and 2-b, 
we show the polarization angle and the degree  
of the emission line from the disk around a black hole.
They are shown as a function of energy for three inclination 
angles of the disk.
The other parameters are 
$R_{{\rm in}}=1.24 M$, $R_{{\rm out}}=20 M$, $q=3$, and $a=0.998 M$.
The inner radius, $R_{{\rm in}}$, is chosen as a marginally stable radius of
the Kepler disk around the rotating black hole with $a=0.998 M.$
Each curve is drawn for the range of the shifted energy.
The angles of these models significantly deviate from 
the value of $ 90^{\circ} $ in a non-relativistic calculation.
The deviation is very clear for a small inclination angle of 
the disk, although the polarization degree is small.
The  degree decreases with the decrease in the inclination angle. 
Indeed, the degree vanishes for the face-on case ($ i=0^{\circ}$), since
the degree is zero for 
the radiation perpendicular to the disk surface, as discussed in the
previous section. 

%3 : Fig.3a-c

In figure 3, we compared the emission from three models in 
the line profile, the polarization angle and the degree.
The three models are different in the inner boundary, 
which is chosen as the radius of the marginally stable orbit
around the black hole. That is, 
$R_{{\rm in}}=6M $ for $ a=0,$  and $R_{{\rm in}}=1.24M, a=0.998M.$  
We also show the case of $R_{{\rm in}}=6M, a=0.998M$
for a comparison.
As shown in previous calculations of the line profiles, 
the effect of black hole rotation 
appears through the inner boundary. 
This is also true for the polarization.
Two models with the same innermost radius, $R_{{\rm in}}=6M $, give
almost the same result, which is consistent with Kojima (1991). 
The effect of the innermost point is clear in only the red-shifted part.

%Fig.4(a)(b)

The results mentioned so far are important for the idealized situation, 
since the spectro-polarimetry of the line is a quite difficult
challenge in future observations. 
We consider the energy-integrated values, which are
useful for an observation without spectro-polarimetry. 
The polarization angle and the degree for 
energy-integrated line photons are shown in figure 4.  
The relativistic effects significantly appear in 
both of polarization angle and the degree for any inclination 
angle.

%Fig.5
Finally, we will apply our model to the broad iron line discovered 
in the Seyfert galaxy MCG-6--30--15. 
From the line profile, some parameters are
inferred (Nandra et al. 1997),
but inconclusive. The present data fit equally well with 
rotating and non-rotating black hole models.
Using their fitting parameters, we 
show the polarization properties 
as well as the line profile in figure 5.
The difference between the rotating and non-rotating models
is evident in the redshift part.
The polarization will indeed give additional information, 
although  the model parameters will not be uniquely determined.

%
% section 4
%
\section{Discussion}

 Chen and Eardley (1991) calculated
the profile and the polarization property of the line
emitted for the mildly relativistic case,
in particular, their calculation is limited to 
a non-rotating black hole. 
We extend their work to the extremely relativistic case
with the rotating black hole, since recent observations suggest 
evidence of  
 relativistic phenomena at a few times the black hole radius.
In such a region of strong gravitational field,
the black hole spin is important, since the inner edge of the
disk depends on it.
Our calculations show that 
the difference in black holes, i.e., spin effect, appears
through the possible range of  emission.
Future observations, despite the difficulties, may produce
 phenomena relevant to our model.

%

%bib
%%%%%%%%%%%%            Bibliography           %%%%%%%%%%%%%%%%%%%
%\clearpage
\vspace{50mm}
\section*{References}
\re 			
Chandrasekhar S.\ 1960, Radiation Transfer(Dover, NY) p24, p234
\re 			
Chen K., Eardley D.M. 1991, ApJ 382, 125
\re 	
Connors P.A., Piran T., Stark R.F. 1980, ApJ 235, 224
\re 	
Dabrowski Y., Fabian A.C., Iwasawa K., Lasenby A.N., Reynolds C.S. 1997,
MNRAS 288, L11 
\re 	
Fabian A.C., Rees M.J., Stella L., White N.E. 1989, MNRAS 238, 729
\re 
Iwasawa K., Fabian A.C., Reynolds C.S., Nandra K., Otani C., Inoue H.,
Hayashida K., Brandt W.N., et al. 1996, MNRAS 282,1038
\re 	
Kojima Y. 1991, MNRAS 250, 629
\re 	
Laor A. 1991, ApJ 376, 90
\re 		
Laor A., Netzer H., Piran T. 1990, MNRAS 242, 560
\re 	
Nandra K., George I.M., Mushotzky R.F., Turner T.J., Yaqoob T. 1997, ApJ
477, 602 
\re 	
Reynolds C.S., Young A.J., Begelman M.C., Fabian A.C. 1999, ApJ 514, 164
\re
Tanaka Y., Nandra K., Fabian A.C., Inoue H., Otani C., Dotani T.,
Hayashida K., Iwasawa K., et al. 1995, Nature 375, 659
\re 	
Weaver K.A., Yaqoob T. 1998,  ApJ 502, L139
	
%
%
%
%\clearpage
%\centerline{Figure Captions}
%\bigskip
%%
%\begin{fv}{1}
%{7cm}
%{Polarization angle emitted from rings as a function of the 
% shifted energy.
% The curves correspond to the radius of the ring, $R=5 M$(solid line),
% $10 M$(dotted line), and  $R=25 M$(dashed line).
% The other parameters are fixed as the inclination angle, $i=45^\circ$,
% and  the black hole spin, $a=0.998 M$.} 
%\end{fv}
\begin{center}
\begin{figure}
\vspace{90mm}
\includegraphics{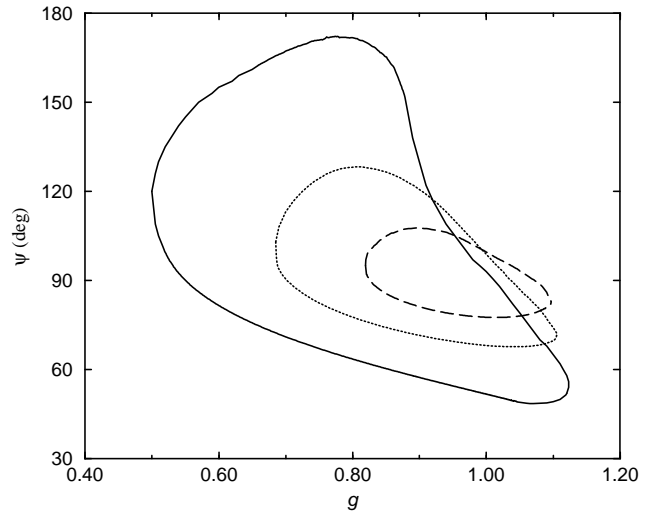}
\caption{Polarization angle emitted from rings as a function of the 
 shifted energy.
 The curves correspond to the radius of the ring, $R=5 M$(solid line),
 $10 M$(dotted line), and  $R=25 M$(dashed line).
 The other parameters are fixed as the inclination angle, $i=45^\circ$,
 and  the black hole spin, $a=0.998 M$.} 
\end{figure}
\end{center}

%%
%\begin{fv}{2}
%{7cm}
%{Effects of the inclination angle:
%({\it a}) the angle and ({\it b}) degree of the polarization.
% Three curves are shown for $i=20^\circ$(solid line),
% $i=45^\circ$(dotted line),  and $i=70^\circ$ (dashed line). 
% The other parameters are chosen as 
% $R_{{\rm in}}=1.24 M$ and $a=0.998 M$.  } 
%\end{fv}
\begin{center}
\begin{figure}
\vspace{125mm}
\includegraphics{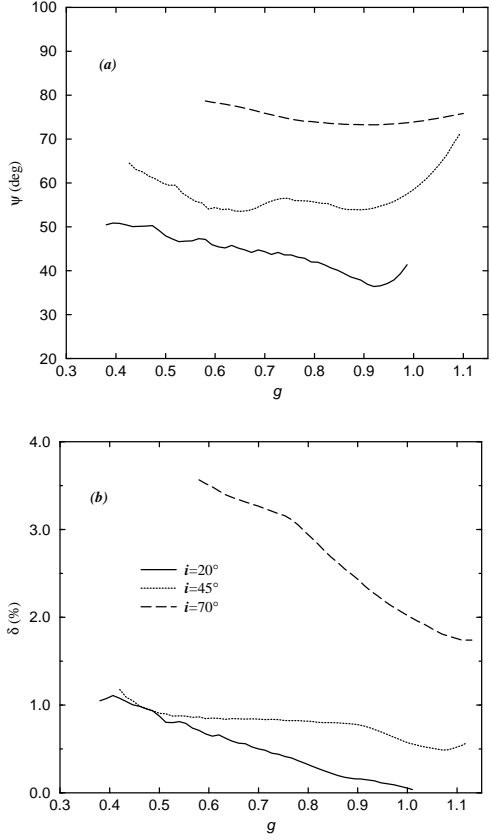}
\caption{Effects of the inclination angle:
({\it a}) the angle and ({\it b}) degree of the polarization.
 Three curves are shown for $i=20^\circ$(solid line),
 $i=45^\circ$(dotted line),  and $i=70^\circ$ (dashed line). 
 The other parameters are chosen as 
 $R_{{\rm in}}=1.24 M$ and $a=0.998 M$.  } 
\end{figure}
\end{center}

%%
%\begin{fv}{3}
%{7cm}
%{Effects of black hole spin on the emission lines:
%({\it a}) line profile, ({\it b}) polarization angle, and ({\it c})
% degree of polarization.  
% The solid line is for the non-rotating case, i.e., $a=0$,
% $R_{{\rm in}}=6M$.  The dashed line is for the rotating case with $a=0.998 M$,
% $R_{{\rm in}}=1.24M$.  We also show  for a comparison the case of 
% $a=0.998 M$, $R_{{\rm in}}=6 M$ by the dotted line.
% The inclination angle is  $i=45^\circ$. } 
%\end{fv}
\begin{center}
\begin{figure}
\vspace{130mm}
\includegraphics{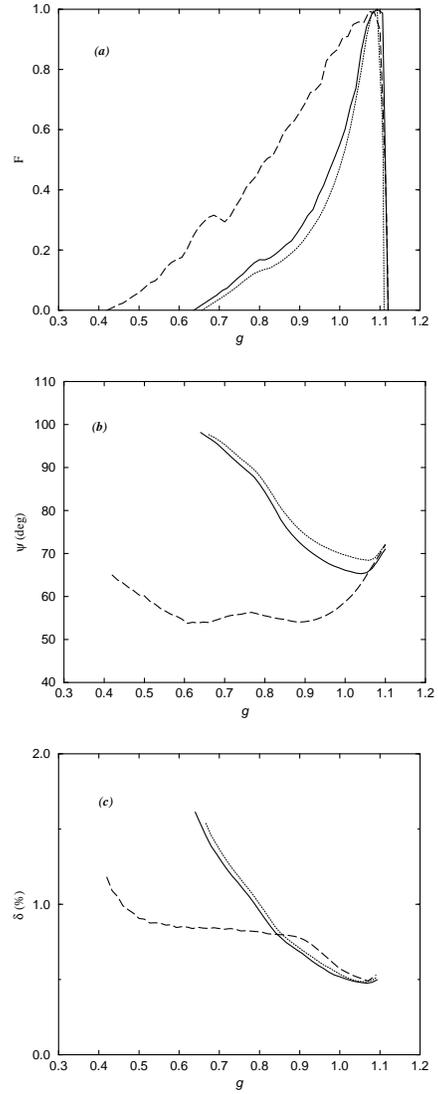}
\caption{Effects of black hole spin on the emission lines:
({\it a}) line profile, ({\it b}) polarization angle, and ({\it c})
 degree of polarization.  
 The solid line is for the non-rotating case, i.e., $a=0$,
 $R_{{\rm in}}=6M$.  The dashed line is for the rotating case with $a=0.998 M$,
 $R_{{\rm in}}=1.24M$.  We also show  for a comparison the case of 
 $a=0.998 M$, $R_{{\rm in}}=6 M$ by the dotted line.
 The inclination angle is  $i=45^\circ$. }  
\end{figure}
\end{center}

%% 
%\begin{fv}{4}
%{7cm}
%{ Polarization as a function of the disk inclination angle 
%  for the energy-integrated light:
%({\it a}) the angle and ({\it b}) degree of the polarization.
% The non-relativistic calculation is shown by the solid line.
% Two models for the relativistic calculation  are also shown.
% The dotted line corresponds to $a=0$ and $R_{{\rm in}}=6M$,
% and the dashed line corresponds to $a=0.998 M$ and $R_{{\rm in}}=1.24M$. 
%}
%\end{fv}
\begin{center}
\begin{figure}
\vspace{170mm}
\includegraphics{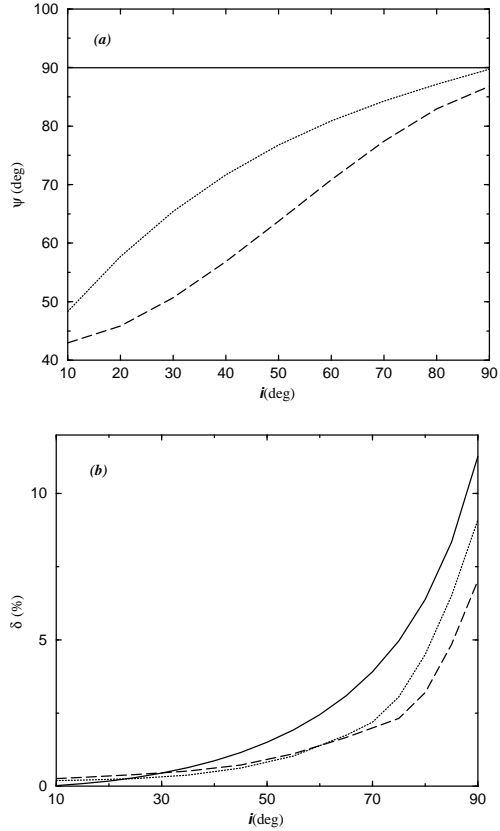}
\caption
{ Polarization as a function of the disk inclination angle 
  for the energy-integrated light:
({\it a}) the angle and ({\it b}) degree of the polarization.
 The non-relativistic calculation is shown by the solid line.
 Two models for the relativistic calculation  are also shown.
 The dotted line corresponds to $a=0$ and $R_{{\rm in}}=6M$,
 and the dashed line corresponds to $a=0.998 M$ and $R_{{\rm in}}=1.24M$. 
} 
\end{figure}
\end{center}

%
%\begin{fv}{5}
%{7cm}
%{Application to the MCG-6--30--15:
%({\it a}) line profile, ({\it b}) polarization angle, ({\it c})
% percentage of polarization of the line.
% The adopted parameters of the solid line are 
% $a=0$,  $R_{{\rm in}}=6M$, $R_{{\rm out}}=20M$,  $q=3$, and $i=34^\circ$.
% The parameters of the dotted line are the same, except for
% $a=0.998 M $ and $R_{{\rm in}}=1.24M$. } 
%\end{fv}
%
\begin{center}
\begin{figure}
\vspace{125mm}
\includegraphics{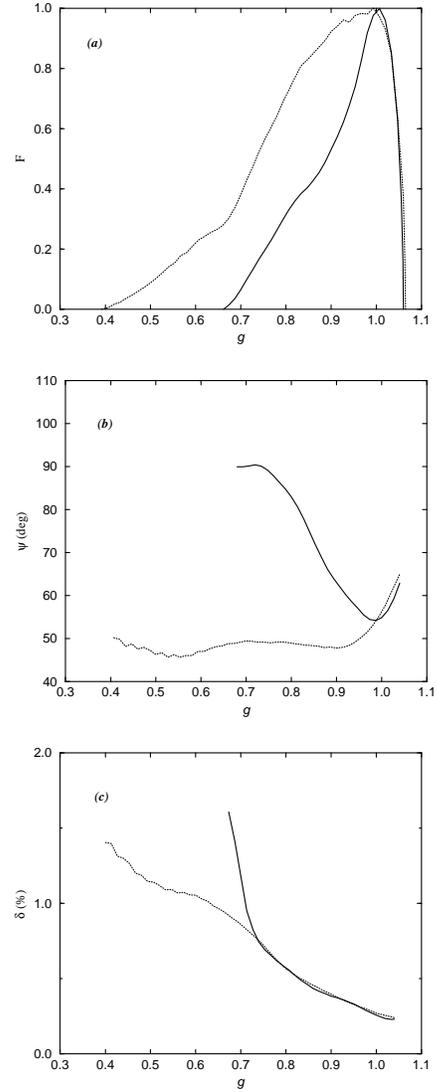}
\caption
{Application to the MCG-6--30--15:
({\it a}) line profile, ({\it b}) polarization angle, ({\it c})
 percentage of polarization of the line.
 The adopted parameters of the solid line are 
 $a=0$,  $R_{{\rm in}}=6M$, $R_{{\rm out}}=20M$,  $q=3$, and $i=34^\circ$.
 The parameters of the dotted line are the same, except for
 $a=0.998 M $ and $R_{{\rm in}}=1.24M$. } 
\end{figure}
\end{center}

\end{document}